\documentclass[a4paper,11pt]{article}
\pdfoutput=1 

\usepackage{jinstpub} 
\usepackage{lineno}

\usepackage[printonlyused]{acronym}
\makeatletter
\AtBeginDocument{%
  \renewcommand*{\AC@hyperlink}[2]{#2}
}
\makeatother

\usepackage{listings}
\usepackage{multirow}
\usepackage{subcaption}
\newcommand*{\doi}[1]{\href{http://doi.org/#1}{doi: #1}}

\graphicspath{{logos/},{graphics/}}

\title{Pixel column issue in the ATLAS Inner Tracker modules}

\author[a,1]{L. Meng,\note{Corresponding author.\\Copyright 2025 CERN for the benefit of the ATLAS Collaboration. Reproduction of this article or parts of it is allowed as specified in the CC-BY-4.0 license.} }

\author[b]{ R. Bates,}
\author[b]{ C. Buttar,}
\author[c]{ G. Calderini,}
\author[c]{ F. Crescioli,}
\author[b]{ L. Cunningham,}
\author[d]{ Y. Dieter,}
\author[b]{ R. Han,}
\author[e]{ T. Heim,}
\author[f]{ S. Hirose,}
\author[d]{ F. Huegging,}
\author[e]{ C. Hultquist,}
\author[g]{ D. Kim,}
\author[h]{ A. Korn,}
\author[i]{ M. Marjanovic,}
\author[g]{ J. Metcalfe,}
\author[h]{ K. Nakamura,}
\author[k]{ J. Pater,}
\author[l]{ H. Pernegger,}
\author[b]{ M.A.A. Samy,}
\author[d]{ M. Schuessler,}
\author[k]{ A. Sharma,}
\author[e]{ E. Thompson,}

\author[m]{ M. Backhaus,}
\author[l]{ J. Christiansen,}
\author[n]{ F. Loddo}

\affiliation[a]{Lancaster University, LA1 4YW Lancaster, United Kingdom}
\affiliation[b]{University of Glasgow, United Kingdom}
\affiliation[c]{LPNHE, Paris, France}
\affiliation[d]{University of Bonn, Germany}
\affiliation[e]{Lawrence Berkeley Laboratory, Berkeley, United States}
\affiliation[f]{University of Tsukuba, Japan}
\affiliation[g]{Argonne National Laboratory, United States}
\affiliation[h]{KEK, Japan}
\affiliation[i]{UCL, London, United Kingdom}
\affiliation[j]{University of Oklahoma, United States}
\affiliation[k]{University of Manchester, United Kingdom}
\affiliation[l]{CERN, Switzerland}
\affiliation[m]{ETH Zuerich, Switzerland}
\affiliation[n]{INFN Sezione di Bari, Italy}

\emailAdd{lingxin.meng@cern.ch}

\abstract{Pixel modules are currently being built for the ATLAS \acs{itk} Pixel detector upgrade.
During the preproduction phase, recurring chip malfunctioning was observed during electrical testing.
It was possible to bypass this issue by disabling some pixel core columns in the ITkPix readout chip.
Therefore the issue is called “core column issue” which is a direct disqualifier for a pixel module.
A concerning number of cases has been observed in pixel modules with ITkPix v1.1 as well as v2 chips which significantly impacts the module yield.
However, the behaviour is erratic and there is not any evidence hinting at the origin of this issue.
These proceedings outline the investigations of the issue, highlighting the electrical behaviour during testing, present findings from the data collected via our production database and through visual inspection, and point towards possible causes of the issue.}


\proceeding{20$^{\text{th}}$ Anniversary "Trento" Workshop on Advanced Silicon Radiation Detectors\\
  4--6 February 2025\\
  FBK, Trento, Italy}

\begin{document}
\maketitle

\flushbottom

\acresetall
\acrodef{adc}[ADC]{analog-to-digital converter}
\acrodef{atlas}[ATLAS]{A Toroidal LHC Apparatus}
\acrodef{asic}[ASIC]{application-specific integrated circuit}
\acrodef{awg}[AWG]{American wire gauge}

\acrodef{bc}[BC]{bunch crossing}
\acrodef{bcid}[BCID]{bunch crossing ID}

\acrodef{cc}[CC]{core column}

\acrodef{cce}[CCE]{charge collection efficiency}
\acrodef{ccpd}[CCPD]{capacitively coupled pixel device}
\acrodef{cern}[CERN]{Conseil Europ\'een pour la Recherche Nucl\'eaire (European Organisation for Nuclear Research)}
\acrodef{ckm}[CKM]{Cabibbo-Kobayashi-Maskawa}
\acrodef{cms}[CMS]{Compact Muon Solenoid}
\acrodef{croc}[CROC]{CMS Readout Chip}

\acrodef{csc}[CSC]{Cathode Strip Chamber}
\acrodef{cv}[C-V]{capacitance-voltage}
\acrodef{cs}[CS]{cluster size}
\acrodef{cpu}[CPU]{central processing unit}

\acrodef{dac}[DAC]{digital-to-analog converter}
\acrodef{daq}[DAQ]{data acquisition}
\acrodef{dcs}[DCS]{detector control system}
\acrodef{dp}[DP]{DisplayPort}
\acrodef{dv}[DV]{design verification}

\acrodef{dut}[DUT]{device under test}

\acrodef{edx}[EDX]{energy-dispersive x-ray}
\acrodef{eh}[e-h]{electron-hole}
\acrodef{enc}[ENC]{equivalent noise charge}

\acrodef{enig}[ENIG]{electroless nickel immersion gold}
\acrodef{enepig}[ENEPIG]{electroless nickel electroless palladium immersion gold}
\acrodef{esd}[ESD]{electrostatic discharge}
\acrodef{ev}[eV]{electron volt}
\acrodef{eoc}[EOC]{end-of-column}
\acrodef{eudet}[EUDET]{European Detector R\&D}

\acrodef{fem}[FEM]{finite element method}
\acrodef{fe}[FE]{front-end}
\acrodef{fpga}[FPGA]{Field-Programmable Gate Array}
\acrodef{fw}[FW]{firmware}
\acrodef{fwhm}[FWHM]{full width at half maximum}
\acrodef{fz}[FZ]{float zone}
\acrodef{cz}[CZ]{Czochralski}

\acrodef{gds}[.gds]{Graphical Design Station II}
\acrodef{gui}[GUI]{graphical user interface}

\acrodef{hitdisc}[HitDiscCnfg]{Hit Discriminator Configuration}
\acrodef{hl}[HL]{High Luminosity}
\acrodef{hpk}[HPK]{Hamamatsu Photonics K.K.}
\acrodef{hv}[HV]{high voltage}
\acrodef{hsio}[HSIO]{high speed input/output}

\acrodef{ibl}[IBL]{Insertable B-Layer}
\acrodef{id}[ID]{Inner Detector}
\acrodef{itk}[ITk]{Inner Tracker}

\acrodef{ifae}[IFAE]{Institut de Fisica d'Altes Energies}

\acrodef{iv}[IV]{sensor leakage current as a function of the bias voltage}

\acrodef{lar}[LAr]{Liquid Argon}
\acrodef{lep}[LEP]{Large Electron-Positron Collider}
\acrodef{lf}[LF]{LFoundry}
\acrodef{lhc}[LHC]{Large Hadron Collider}
\acrodef{lp}[LP]{low power}
\acrodef{lpgbt}[lpGBT]{Low Power GigaBit Transceiver}
\acrodef{ls}[LS]{Long Shutdown}
\acrodef{lsb}[LSB]{least significant bit}
\acrodef{lv}[LV]{low voltage}

\acrodef{maps}[MAPS]{monolithic active pixel sensor}
\acrodef{met}[MET]{missing transverse energy}
\acrodef{mdt}[MDT]{Monitored Drift Tube}
\acrodef{mip}[MIP]{minimum ionising particle}
\acrodef{mos}[MOS]{metal-oxide-semiconductor}
\acrodef{mosfet}[MOSFET]{metal-oxide-semiconductor field-effect transistor}
\acrodef{mpv}[MPV]{most probable value}
\acrodef{mwpc}[MWPC]{multi-wire proportional chamber}
\acrodef{mpw}[MPW]{multi-project wafer}
\acrodef{mux}[MUX]{multiplexer}

\acrodef{niel}[NIEL]{non-ionising energy loss}
\acrodef{neq}[neq]{1\,MeV neutron equivalent fluence}
\acrodef{nfe}[NFE]{nearly-free electron}
\acrodef{ntc}[NTC]{negative temperature coefficient}

\acrodef{opamp}[op-amp]{operational amplifier}
\acrodef{os}[OS]{operating system}

\acrodef{ovp}[OVP]{over-voltage protection}
\acrodef{usp}[USP]{under-shunt protection}

\acrodef{pcb}[PCB]{printed circuit board}
\acrodef{pdb}[PDB]{production database}
\acrodef{pdk}[PDK]{process design kit}
\acrodef{pka}[PKA]{primary knock-on atom}
\acrodef{pmt}[PMT]{photomultiplier tube}

\acrodef{ppd}[PPD]{pixelized photodiode}
\acrodef{preamp}[preamp]{preamplifier}
\acrodef{psd}[PSD]{power spectral density}
\acrodef{psu}[PSU]{power supply unit}

\acrodef{ps}[PS]{power supply}

\acrodef{qa}[QA]{quality assurance}
\acrodef{qc}[QC]{quality control}

\acrodef{raf}[RAF]{rise-and-flatten}
\acrodef{rce}[RCE]{Reconfigurable Cluster Element}
\acrodef{rf}[RF]{radio frequency}
\acrodef{rod}[ROD]{Readout Driver}
\acrodef{roi}[RoI]{region of interest}
\acrodef{roc}[ROC]{readout cell}
\acrodef{rock}[RoCk]{readout clock}
\acrodef{rpc}[RPC]{Resistive Plate Chamber}
\acrodef{rms}[RMS]{root mean square}

\acrodef{set}[SET]{Smart Equipment Technology}
\acrodef{sct}[SCT]{Semiconductor Tracker}
\acrodef{sldo}[SLDO]{Shunt Low Drop Out regulator}
\acrodef{smd}[SMD]{surface-mount device}
\acrodef{syres}[SyRes]{synchronous reset}
\acrodef{scifi}[SciFi]{scintillating fiber}
\acrodef{sio}[SiO\textsubscript{2}]{silicon oxide}
\acrodef{sipm}[SiPM]{silicon photomultiplier}
\acrodef{slac}[SLAC]{Stanford Linear Accelerator Center}
\acrodef{sn}[SN]{serial number}
\acrodef{sps}[SPS]{Super Proton Synchrotron}

\acrodef{tpg}[TPG]{thermal pyrolytic graphite}
\acrodef{tdr}[TDR]{technical design report}
\acrodef{tid}[TID]{total ionising dose}
\acrodef{tim}[TIM]{thermal interface material}
\acrodef{tw}[TW]{time walk}
\acrodef{tot}[ToT]{time-over-threshold}
\acrodef{trt}[TRT]{Transition Radiation Tracker}
\acrodef{ttl}[TTL]{transistor-transistor logic}

\acrodef{tof}[ToF]{time-of-flight}

\acrodef{ups}[UPS]{uninterruptable power supply}
\acrodef{uv}[UV]{ultra violet}

\acrodef{vi}[VI]{input voltage Vin as a function of the input current Iin for \acs{sldo}}
\acrodef{vhdl}[VHDL]{Very High Speed Integrated Circuit Hardware Description Language}
\acrodef{xml}[XML]{Extensible Markup Language}

\acrodef{zif}[ZIF]{zero insertion force}

\section{Introduction}
\label{sec:intro}

\begin{figure}[h]
\centering
\begin{subfigure}[b]{0.29\textwidth}
\includegraphics[width=\textwidth]{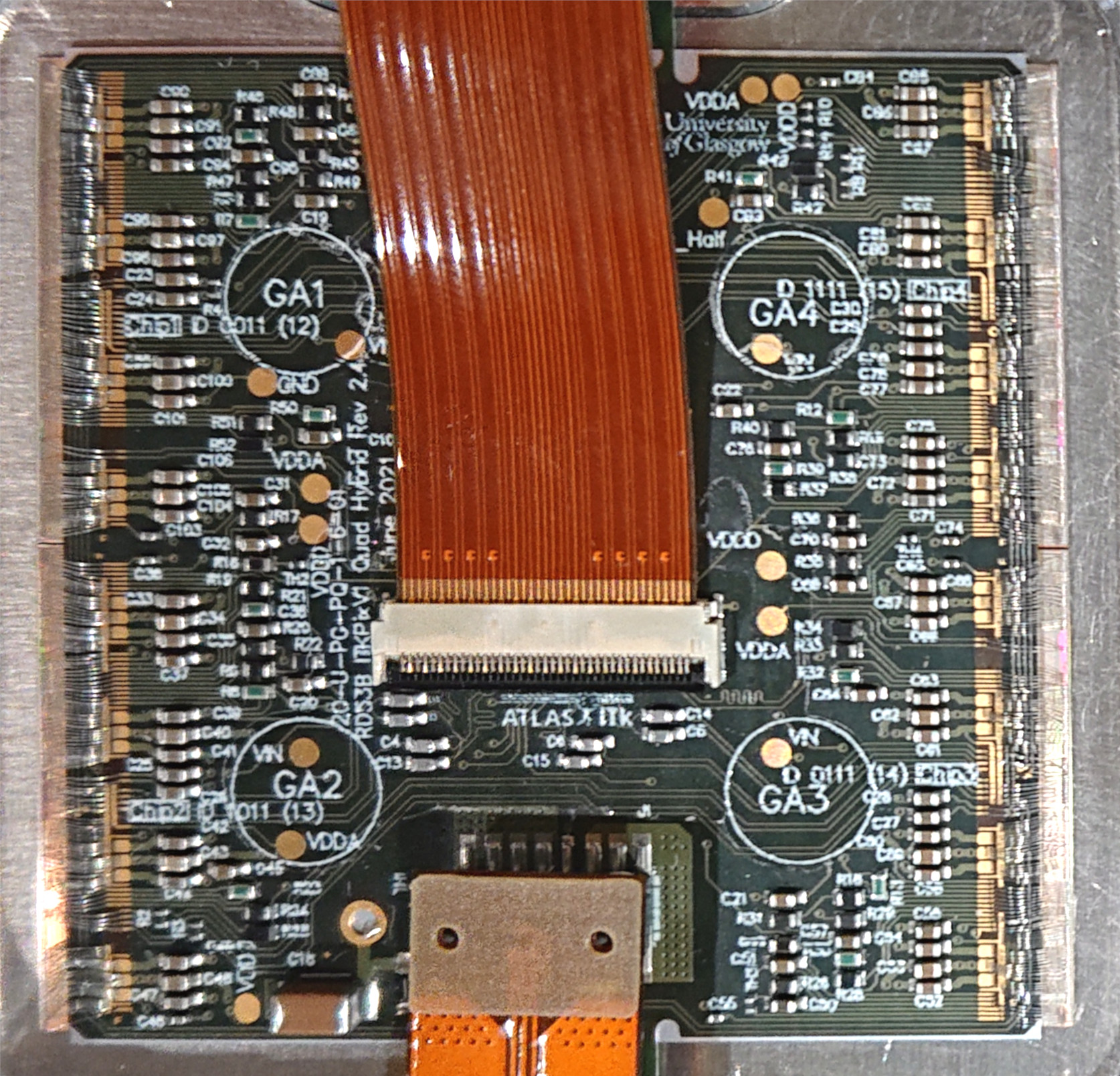}
\caption{}
\label{fig:quad}
\end{subfigure}
\hspace{0.05\textwidth}
\begin{subfigure}[b]{0.39\textwidth}
\includegraphics[width=\textwidth]{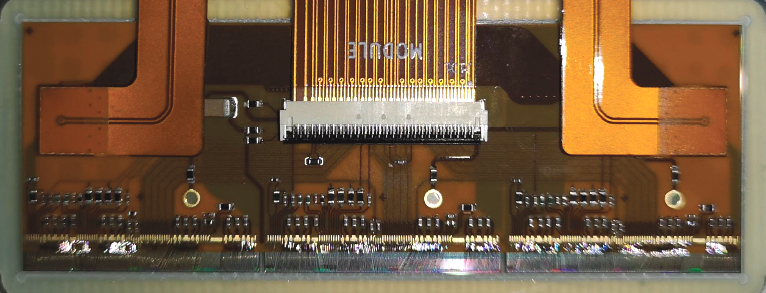}
\caption{}
\label{fig:triplet}
\end{subfigure}

\caption{Two pixel module flavours: (a) a quad module, (b) a linear triplet module.}
\label{fig:quad-triplet-module}
\end{figure}

The \ac{itk} is the upgrade tracker for ATLAS \cite{atlas} in the \ac{hl}-\ac{lhc} era.
The pixel detector \cite{itk} is currently in its production phase, with more than 10000 hybrid modules being built.
The modules come in two flavours: quad modules (figure \ref{fig:quad}) consisting of four \ac{fe} chips and a quad-size planar sensor tile of about $4\times4$\,cm$^2$, and triplet modules of different geometries, figure \ref{fig:triplet}, consisting of three bare modules, each made of one \ac{fe} chip and one single 3D sensor tile.
In the \ac{itk} pixel detector, about 95\% of the modules are quad modules while 5\% are triplet modules.

\begin{figure}
\centering
\begin{subfigure}[b]{0.66\textwidth}
\includegraphics[width=\textwidth]{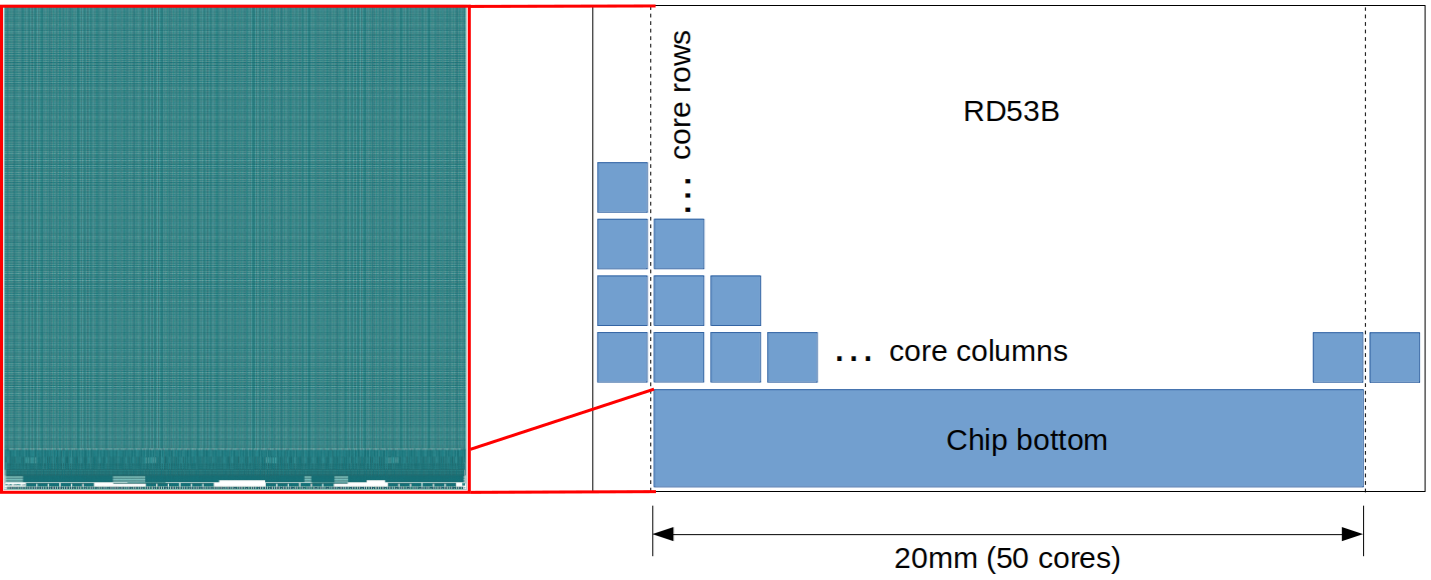}
\caption{}
\label{fig:chipcc}
\end{subfigure}
\hfill
\begin{subfigure}[b]{0.33\textwidth}
\includegraphics[width=\textwidth]{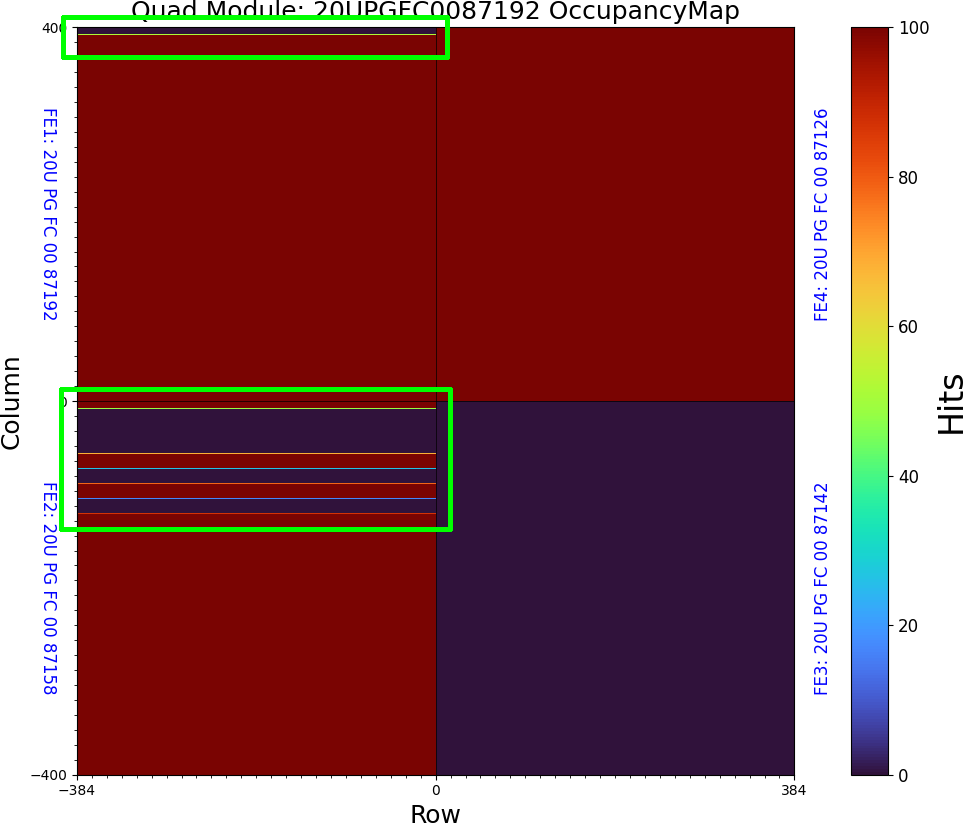}
\caption{}
\label{fig:digitalscan}
\end{subfigure}

\caption{(a) The ITkPix chip layout: pixel matrix (left) with \acl{cc} logic (right). (b) Occupancy map from a digital scan with 100 injections on a module with some \aclp{cc} disabled (marked in green). The bottom right chip was not wirebonded.}
\label{fig:chip-core-columns}
\end{figure}

The \ac{fe} chip was developed by the RD53 collaboration \cite{rd53} for both the ATLAS and CMS inner tracker upgrade for the \ac{hl}-\ac{lhc}.
They differ only in the size of the matrix and in the analog pixel circuit.
The ATLAS variant is called "ITkPix" and consists of 400 columns $\times$ 384 rows, resulting in a total of 153600 pixels.
The digital logic is grouped into 50 \acp{cc} as indicated in figure \ref{fig:chipcc}, where each \acl{cc} consists of 8 pixel columns.

\section{Core Column Issue}

Throughout module preproduction with ITkPix v1.1 and during production using ITkPix v2, certain chip malfunctioning is observed on chips classified as fully functionining - "green" - by wafer probing.
The symptoms are erratic and not consistent, e.g. chip readout is stuck, or chip sends continuous bad data etc.
Chip operation can be recovered by disabling one or more \aclp{cc} (figure \ref{fig:digitalscan}) so that the chip behaves normally.
Thus this kind of chip malfunctioning is called "core column issue".
Because a \acl{cc} contains 2\% of the pixels on the chip, disabling one core column would directly fail a module by our current \ac{qc} criteria (total $<$1\% bad pixels).
This issue is observed on ITkPix v1.1 modules and ITkPix v2 modules, as well as by CMS on their \ac{croc} v1 modules.

\subsection{Statistics}

To obtain an overview of the frequency with which the core column issue occurs, queries were made to the \ac{itk} \ac{pdb} \cite{pdb} which records information of all components, their stages and tests.
\ac{fe} chip configurations were queried to obtain the settings of the core-column-enable registers which contain information about disabled core columns.
A snapshot of the \ac{cc} statistics is shown in table \ref{tab:stats}.
The data are sorted by chip version and testing temperature in order to show any systematics. 

\begin{table}[h]
\caption{
Data extracted from the production database.
The columns "CC/chips(modules)" are fractions of chip (modules) with at least one disabled \acp{cc} out of the total number of chips (modules).
The testing temperatures "warm" and "cold" correspond to about 20\,$^\circ$C and $-15\,^\circ$C, respectively.
}

\begin{tabular}{|c|c|r|r|r|r|}
\hline
chip version               & testing temperature & \multicolumn{1}{c|}{CC/modules} & \multicolumn{1}{c|}{CC/modules (\%)} & \multicolumn{1}{c|}{CC/chips} & \multicolumn{1}{c|}{CC/chips (\%)} \\ \hline
\multirow{2}{*}{v1.1} & warm & 77/357                       & 21.6                              & 116/1393                   & 8.3                             \\ \cline{2-6} 
                      & cold & 42/249                       & 16.9                              & 57/982                     & 5.8                             \\ \hline
\multirow{2}{*}{v2}   & warm & 17/118                       & 14.4                              & 25/470                     & 5.3                             \\ \cline{2-6} 
                      & cold & 9/85                         & 10.6                              & 12/340                     & 3.5                             \\ \hline
\end{tabular}
\label{tab:stats}
\end{table}

\begin{figure}

\centering
\begin{subfigure}[b]{0.32\textwidth}
\includegraphics[width=\textwidth]{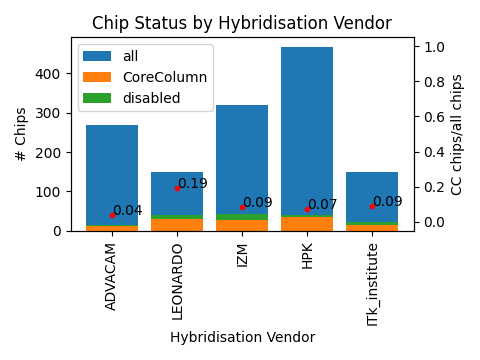}
\caption{}
\label{fig:hvendorchip}
\end{subfigure}
\begin{subfigure}[b]{0.32\textwidth}
\includegraphics[width=\textwidth]{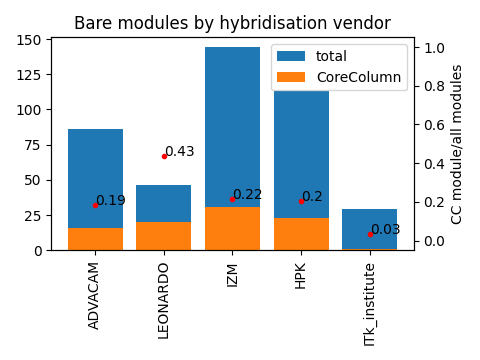}
\caption{}
\label{fig:hvendormodule}
\end{subfigure}
\begin{subfigure}[b]{0.32\textwidth}
\includegraphics[width=\textwidth]{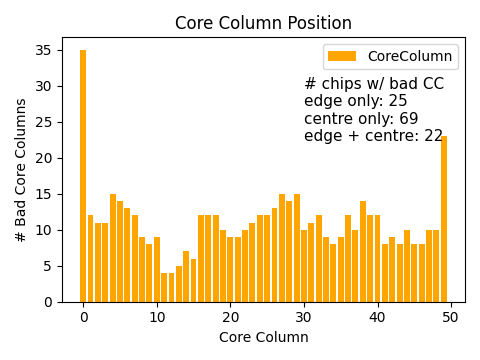}
\caption{}
\label{fig:ccposition}
\end{subfigure}

\caption{Fraction of chips (a) and modules (b) with disabled core columns per hybridisation vendor.
(c) positions of disabled \acp{cc} on all problematic chips.}
\label{fig:hybridisationvendor}
\end{figure}

The largest amount of the data are from warm tests on v1.1 chips assembled in both quad and triplet modules, therefore they are used in the following investigations.
Figure \ref{fig:hybridisationvendor} shows fractions of chips (\ref{fig:hvendorchip}) or modules (\ref{fig:hvendormodule}) with \ac{cc} issues, which are similar across all hybridisation vendors, showing that the issues are not associated with one vendor.
The number of \ac{fe} chips that are disabled in software during testing is also indicated in figure \ref{fig:hvendorchip}.
This could be due to either the user not associating errors with the core column issue, or other issues such as the chip is not wirebonded due to damage of the wirebond pads or wirebonding failure.
Therefore, disabled chips are not included in the number of \ac{fe} chips with \ac{cc}.

\begin{figure}
\centering

\begin{subfigure}[b]{0.49\textwidth}
\includegraphics[width=\textwidth]{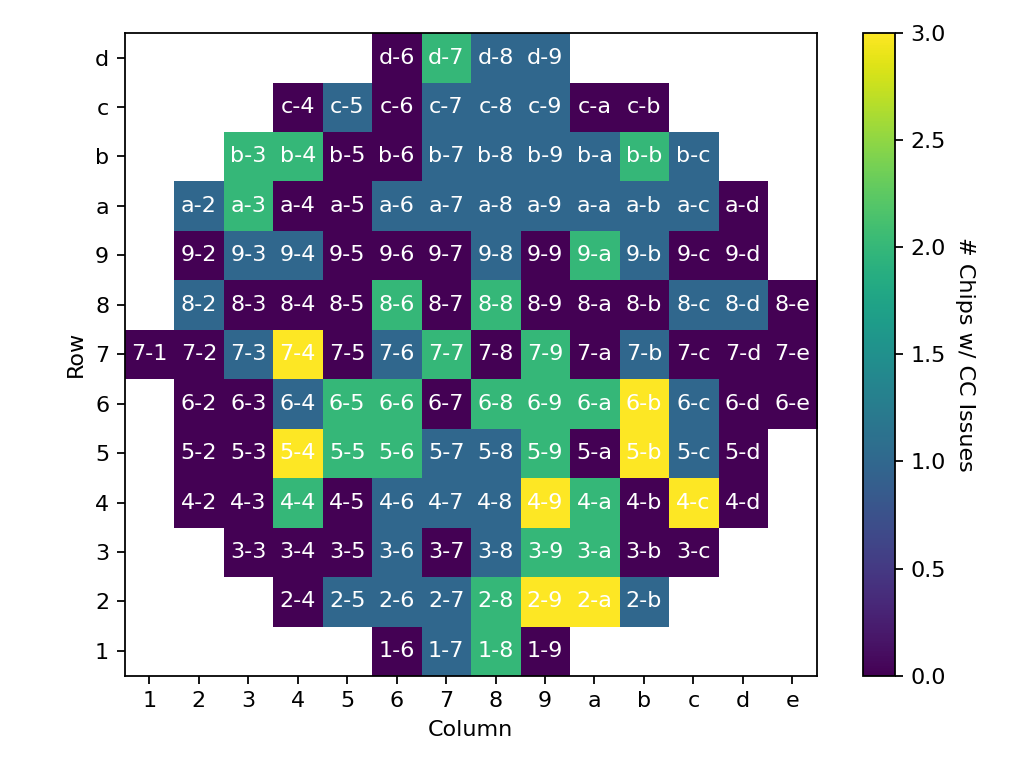}
\caption{}
\label{fig:wafermap}
\end{subfigure}
\hfill
\begin{subfigure}[b]{0.49\textwidth}
\includegraphics[width=\textwidth]{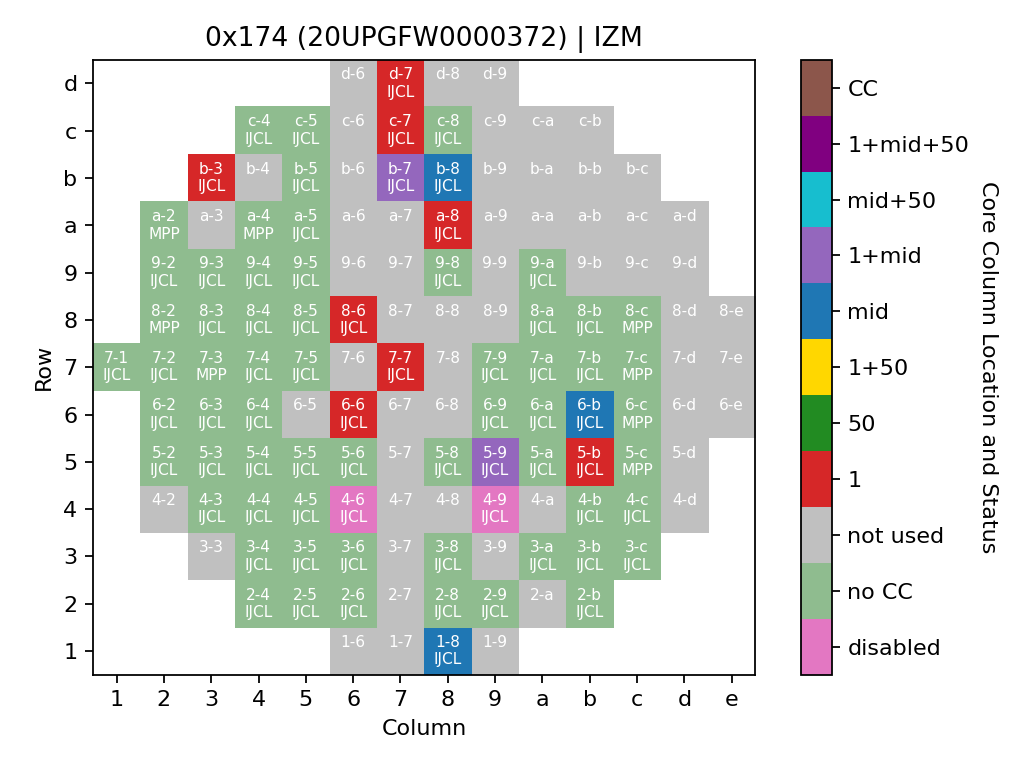}
\caption{}
\label{fig:0x174}
\end{subfigure}

\caption{(a) Cumulative wafer map with occurence of \ac{cc} chip position. (b) example wafer map where \ac{cc} chips often appear in consecutive oder in a row or column. The z-scale indicate the location of disabled core columns on that chip.}
\label{fig:wafermaps}
\end{figure}

Figure \ref{fig:ccposition} shows the distribution of all disabled \acp{cc} for all hybridisation vendors.
The distribution is uniform across \ac{fe} chip except for the edges where an increase is seen.
The increase in disabled edge columns is associated with one vendor.
The position of chips on wafers were studied as shown in figure \ref{fig:wafermaps}.
While the cumulative map (figure \ref{fig:wafermap}) does not show any preferencial position for \ac{cc} chips, looking at individual wafer maps reveals that \ac{cc} chips often cluster together in a row or in a column which could indicate problems with dicing or picking, like shown in figure \ref{fig:0x174}.

Other variables, such as the assembly or testing date or site, sensor leakage current etc, showed no relation to the core column issue. 

\section{Investigations}

Intrisic issues of the design or the chip itself can be excluded to be origins of the core column issue because such would have been identified, flagged or rejected accordingly during the probing of the \ac{fe} chip wafers.
The only stages at which changes happen to the chips are between wafer probing and initial testing, which would be hybridisation and module assembly.
In the hybridisation stage, chips are bump-deposited, thinned, diced and flip-chipped into a bare module.
The bare module is glued onto a flexible module \ac{pcb} and wirebonded in the assembly stage.
Especially hybridisation is a delicate process during which mechanical damages can occur.

\subsection{Probing of diced chips}
\label{sec:single-chip-probing}


To test this hypothesis, 63 and 25 diced ITkPix v2 chips from two different wafers were probed. While the former were all probed to be "green" on the wafer level, the latter were not probed prior to dicing.
Only four from the originally green chips had any pixel issues, which are still categorised as "green" after probing.
Two of them had bad double-column bias to the analog \ac{fe} circuit, and the other two chips likely had \ac{daq} instability that showed up in the scan mask pattern.
None of them showed \ac{cc} issues based on what has been observed.

\subsection{Inconsistencies and analog dependencies}

Bad core columns often manifest themselves already at register reading or in digital scans, but also analog dependencies have been observed.

\begin{figure}
\centering

\begin{subfigure}[b]{0.3\textwidth}
\includegraphics[width=\textwidth]{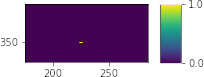}
\caption{}
\label{fig:pixelregion}
\end{subfigure}
\begin{subfigure}[b]{0.3\textwidth}
\includegraphics[width=\textwidth]{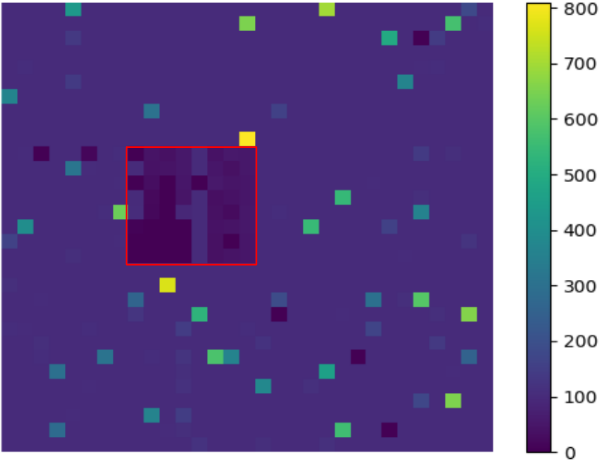}
\caption{}
\label{fig:pixelcore}
\end{subfigure}

\begin{subfigure}[b]{0.4\textwidth}
\includegraphics[width=\textwidth]{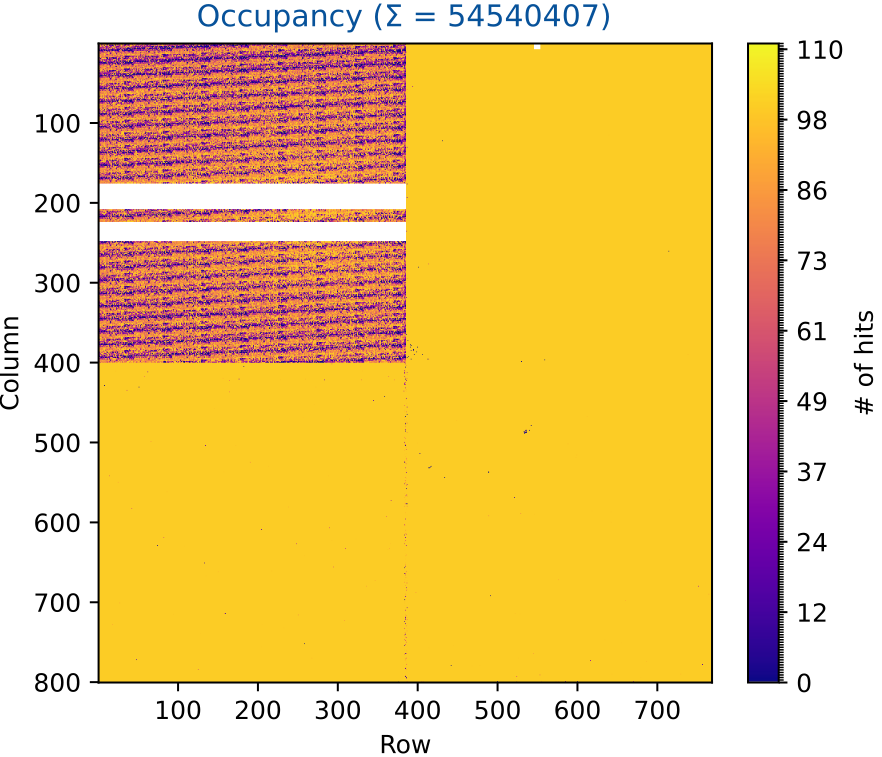}
\caption{}
\label{fig:1650e}
\end{subfigure}
\begin{subfigure}[b]{0.4\textwidth}
\includegraphics[width=\textwidth]{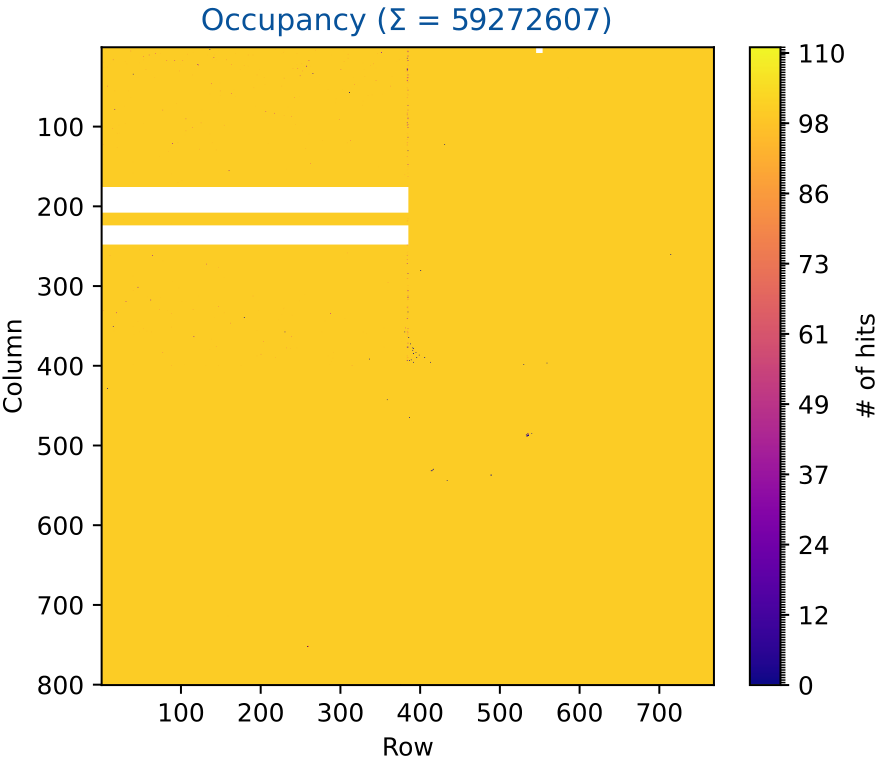}
\caption{}
\label{fig:2600e}
\end{subfigure}

\caption{(a) Map from a pixel test highlighting a bad 4$\times$1 pixel region.
(b) Occupancy map from an analog scan with a bad 8$\times$8 pixel core with mostly 0 hits.
(c) Occupancy map of a module at 1650\,e injection charge.
(d) Occupancy map of the same module at 2600\,e injection charge.
}
\label{fig:scans}
\end{figure}

Very few \ac{cc} cases can be potentially narrowed down to bad pixel regions (4$\times$1 pixels) as shown in figure \ref{fig:pixelregion} or bad pixel cores (8$\times$8 pixels) as shown in figure \ref{fig:pixelcore}.
Once these smaller areas are disabled, the chip can function normally without disabling the \acp{cc}.
This would drastically reduce the number of bad pixels overall on a chip compared to disabling entire core columns.

\ac{cc} issue often shows inconsistent symptoms even within the same chip.
One example is figure \ref{fig:pixelcore}, where the bad pixel core was hard to find.
While the software diagnostics tool cannot find any bad \acp{cc} in this chip, the analog scan often fails, indicating \ac{cc} issues.
Only when sometimes an analog scan runs successfully, this bad pixel core can be seen in the occupancy map.

\ac{cc} symptoms sometimes show dependencies on the analog settings, for example threshold or injection charge.
Figure \ref{fig:1650e} shows a bad occupancy map of the chip with \ac{cc} issue when the injected charge is 1650\,e.
The same chip behaves normally when the injected charge is 2600\,e as shown in figure \ref{fig:2600e}.

When a chip gets stuck due to bad \acp{cc}, a reset signal must be issued, which is not the normal mode of operation.
Modules with continuous data output were studied to confirm that these data cannot be decoded.

\subsection{Mechanical damage}

Core column issues were observed on badly diced chips.
Therefore, we tried to obtain a correlation with mechanical damage.

\paragraph{First hint}

\begin{figure}
\begin{center}
\includegraphics[width=0.75\textwidth]{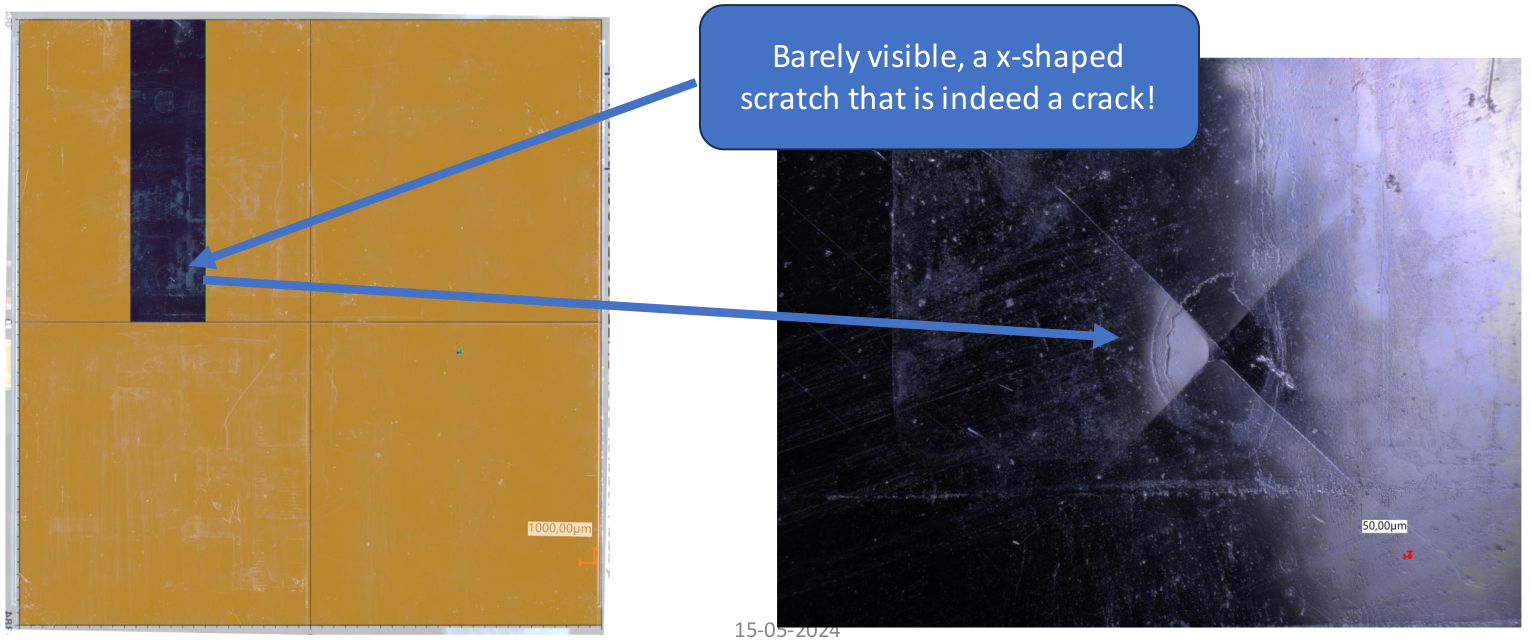}
\end{center}
\caption{A quad module has developed core column issue. Careful visual inspection revealed a crack within the area of disabled core columns which came from a solder splash on the vacuum chuck.}
\label{fig:crack}
\end{figure}

The first confirmation of mechanical damage causing the \ac{cc} issue was an accident in which two previously functional modules developed \ac{cc} issues.
On one module the \ac{cc} issue was reproduced in a different chip by rotating the module by 180$^\circ$ in the same setup.
Only under certain angle and with the correct lighting conditions, some cross or pyramid pattern was found on the back side of the \ac{fe} chips, which could be a crack in the silicon as can be seen in figure \ref{fig:crack}.
The location of this crack lies in the area of the disabled \acp{cc} on the chip.
Investigations found the cause to be a bump from a solder splash of about $100\,\upmu$m size on the vacuum chuck used in that particular setup.

\paragraph{Visual inspection}

\begin{figure}

\begin{subfigure}[b]{0.43\textwidth}
\includegraphics[height=15mm]{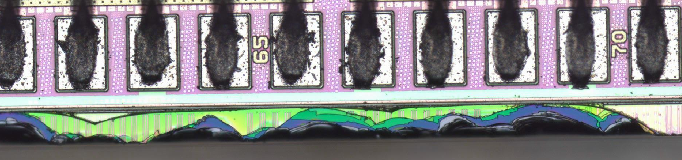}
\caption{}
\label{fig:badedge}
\end{subfigure}
\hfill
\begin{subfigure}[b]{0.56\textwidth}
\includegraphics[height=15mm]{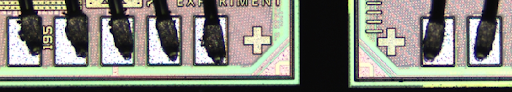}
\caption{}
\label{fig:goodedge}
\end{subfigure}

\caption{Different quality of dicing edges in v1.1 and v2 chips. Core column issues have been observed both on chips with bad (a) and good (b) dicing edges.}
\label{fig:dicing}
\end{figure}

Visual inspection of many affected modules were done.
While a few known chips with bad dicing edges, as shown in figure \ref{fig:badedge}, had core column issues, most \ac{cc} modules collected for this campaign did not show any vislble bad dicing as shown in figure \ref{fig:goodedge}, or any damage on accessible part of the chip, e.g. the backside or around the wirebond pads.
There was no clear correlation between modules with \ac{cc} issue due to bad dicing.
Bad \ac{cc} issues can be associated with poor dicing, but poor dicing does not necessarily imply \ac{cc} issue.

\paragraph{Induce surface damage}

\begin{figure}
\centering

\begin{subfigure}[b]{0.4\textwidth}
\includegraphics[width=\textwidth]{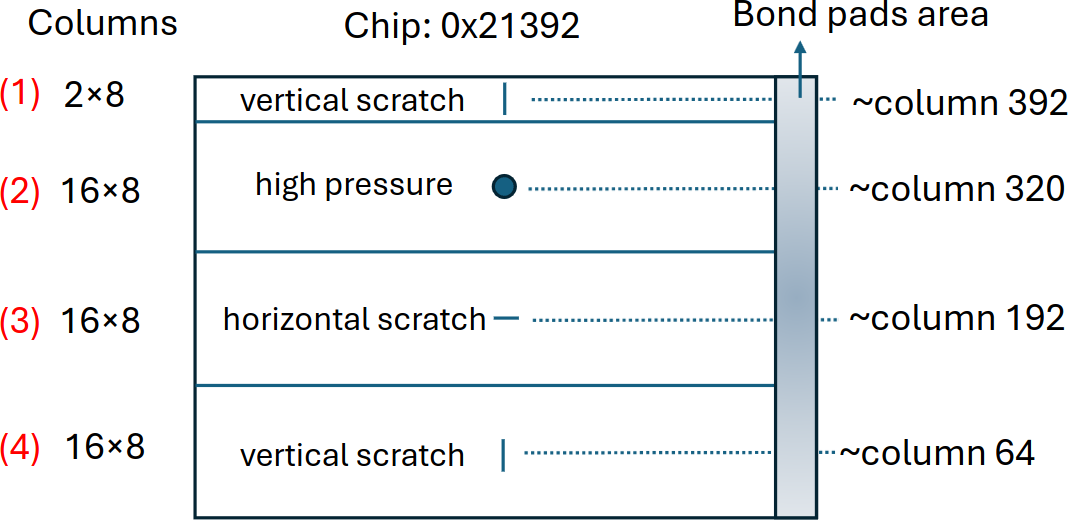}
\caption{}
\label{fig:scratchpattern}
\end{subfigure}
\begin{subfigure}[b]{0.3\textwidth}
\includegraphics[width=\textwidth]{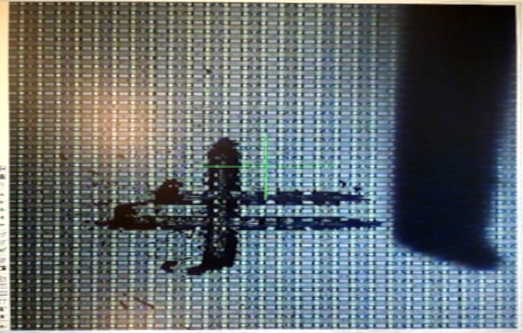}
\caption{}
\label{fig:scratchmark}
\end{subfigure}
\begin{subfigure}[b]{0.27\textwidth}
\includegraphics[width=\textwidth]{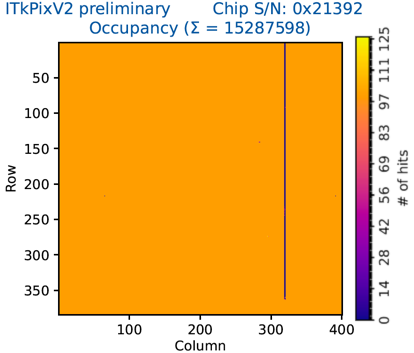}
\caption{}
\label{fig:double-column}
\end{subfigure}

\caption{Destructive test on diced v2 chips by inducing different scratch patterns in different regions of a chip (a) and (b).
The result was a chip with partially damaged double-column analog bias that is visible in an analog scan (c).}
\label{fig:scratch}
\end{figure}

From the diced single ITkPix v2 chips in section \ref{sec:single-chip-probing}, mechanical damage was induced trying to provoke core column issue.
Each chip is divided into several sections as shown in figure \ref{fig:scratchpattern}.
In each of the sections, a different scratch pattern with varying strength was induced via probe needles.
An example of a scratch is shown in figure \ref{fig:scratchmark}.
The chips were then probed again to test for any induced damage.
From the two scratched chips, one ended up dead, the other one suffered a damaged double-column analog bias from one of the scratches, as can be seen in an analog scan in figure \ref{fig:double-column}.
This agrees with expectations as the power distribution of the chip happens in the top metal layers.
However, core column issues were not induced by this scratching test.

\paragraph{Rejected chips}

\begin{figure}
\includegraphics[height=30mm]{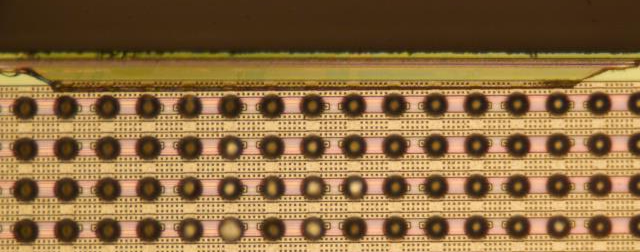}\hfill
\includegraphics[height=30mm]{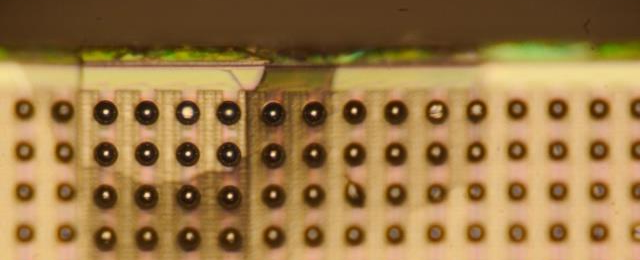}

\vspace{1mm}

\includegraphics[width=0.24\textwidth]{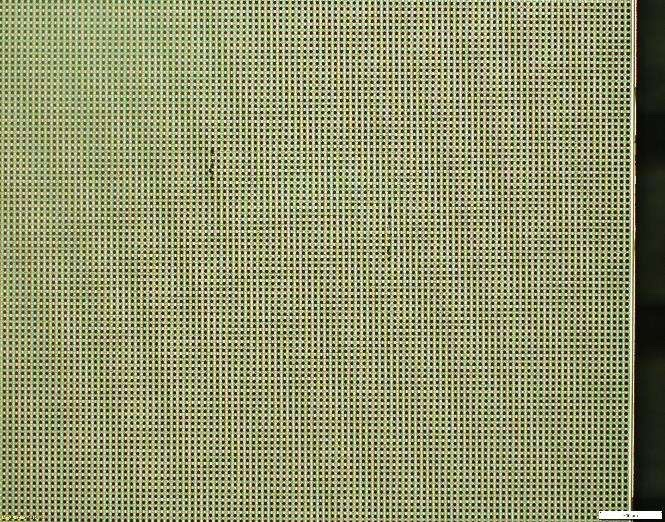}\hspace{0.013\textwidth}\includegraphics[width=0.24\textwidth]{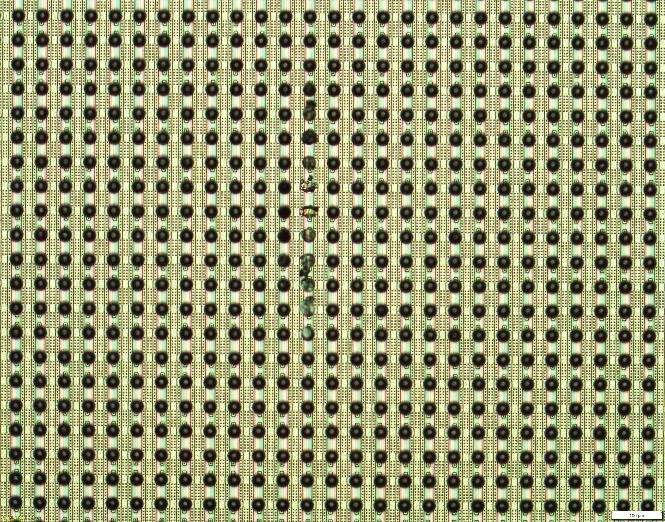}\hspace{0.012\textwidth}
\includegraphics[width=0.24\textwidth]{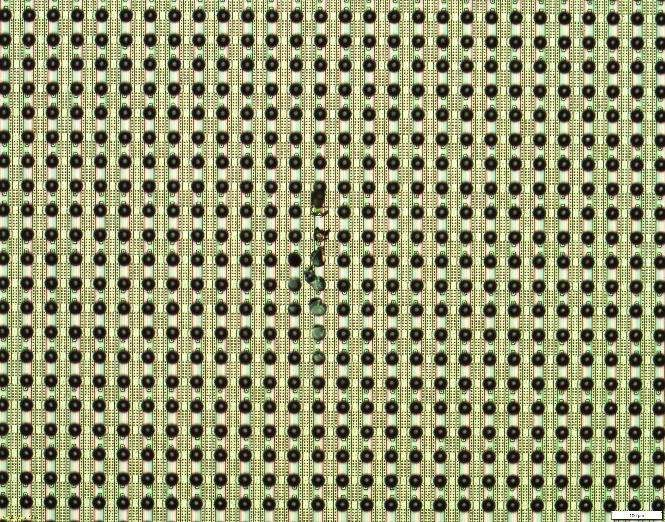}\hspace{0.01\textwidth}\includegraphics[width=0.24\textwidth]{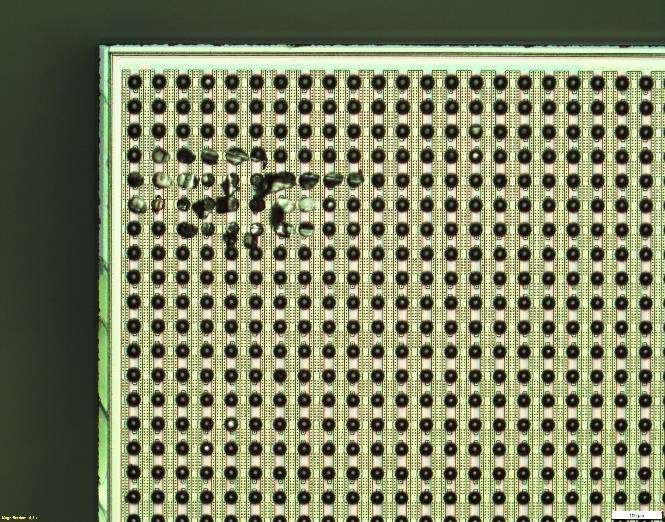}

\caption{Top: damages of two vendor-rejected chips which exhibit core column issue.
Bottom: defects of four vendor-rejected chips without core column issue.}
\label{fig:rejected-chips}
\end{figure}

Chip defects were studied that were grounds for rejection by hybridisation vendors.
Six vendor-rejected chips with different defects were collected, visually inspected with a higher magnification and tested on a probe station.
Using the same software to test all these chips, the two chips with what looks like chipped edges as shown in the top row in figure \ref{fig:rejected-chips} had \ac{cc} issues while no core column issues were found on the other four chips with bump defects as shown in the bottom row in figure \ref{fig:rejected-chips}.

\begin{figure}

\label{fig:rejected-chips-nocc}

\end{figure}

\paragraph{Destructive tests}

\begin{figure}

\begin{subfigure}[b]{0.38\textwidth}
\includegraphics[height=45mm]{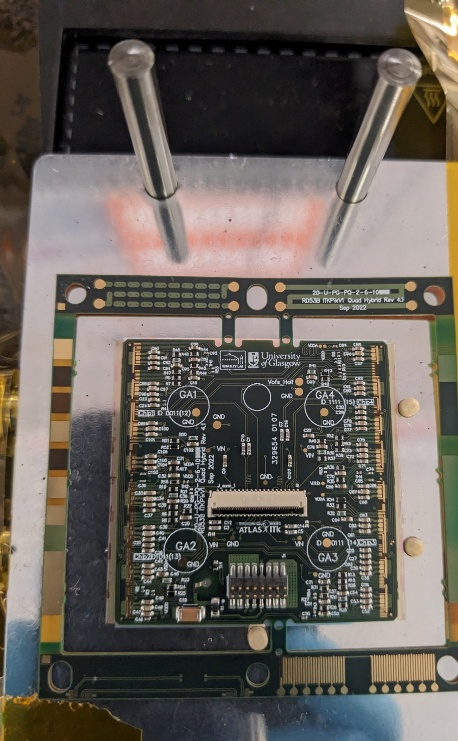}\includegraphics[height=45mm]{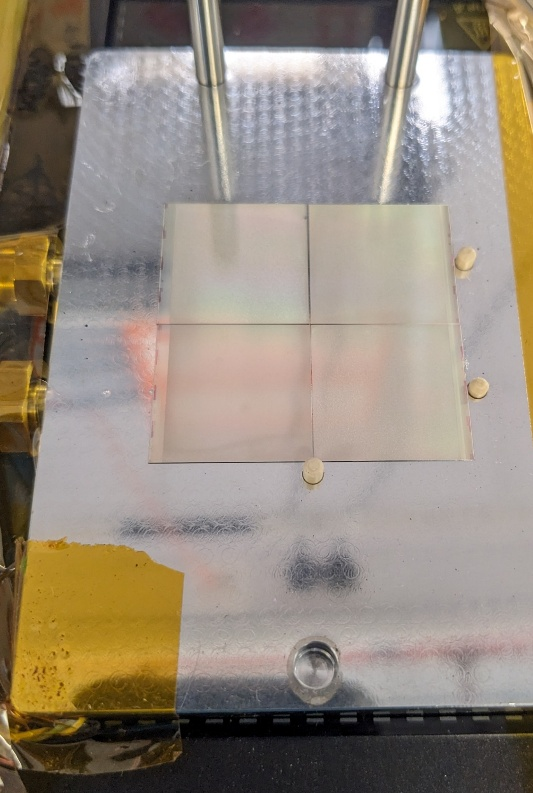}
\caption{}
\label{fig:hotplate}
\end{subfigure}
\hspace{0.002\textwidth}
\begin{subfigure}[b]{0.3\textwidth}
\includegraphics[height=45mm]{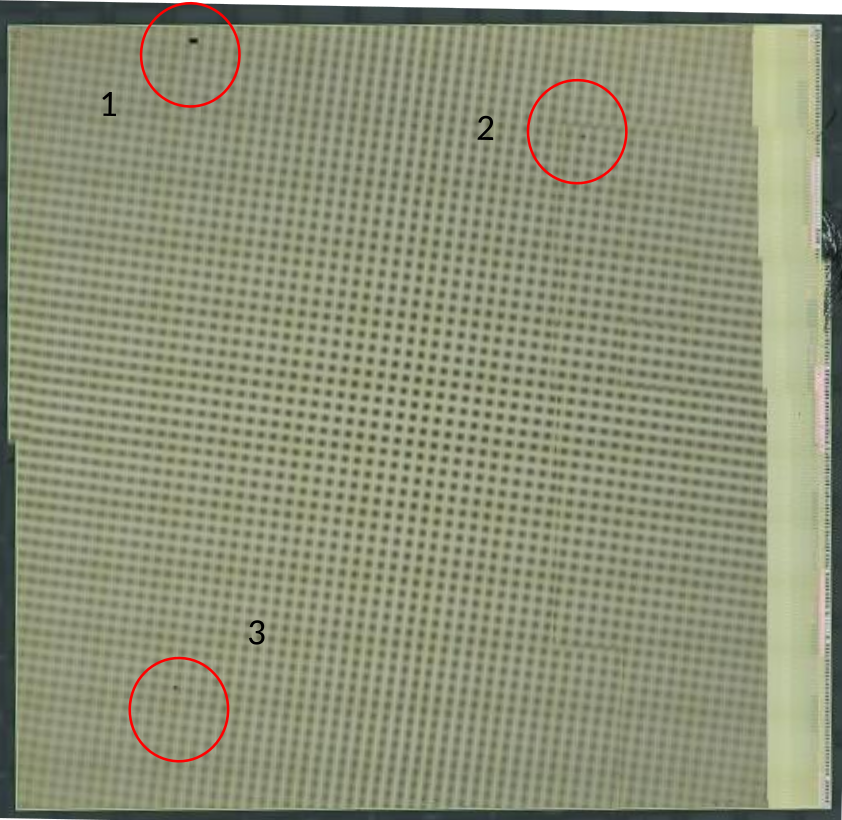}
\caption{}
\label{fig:chip3defects}
\end{subfigure}
\hspace{0.002\textwidth}
\begin{subfigure}[b]{0.3\textwidth}
\includegraphics[height=45mm]{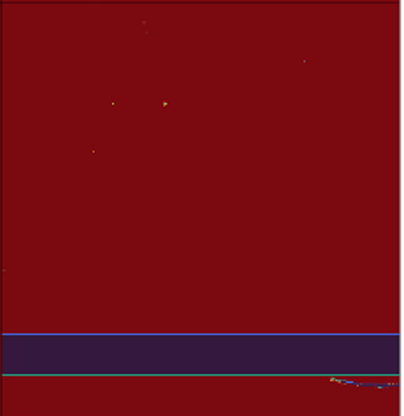}
\caption{}
\label{fig:chip3scan}
\end{subfigure}

\caption{(a) A module on a hotplate (left). After successful disassembly, four chips can be seen on the hotplate (right).
(b) Defects found on one disassembled chip. Out of the three defects, the location of defect 3 agrees with the location of the disabled \acp{cc} in a scan (c).
}
\label{fig:disassembly}
\end{figure}

\begin{figure}

\begin{subfigure}[b]{0.29\textwidth}
\includegraphics[width=\textwidth]{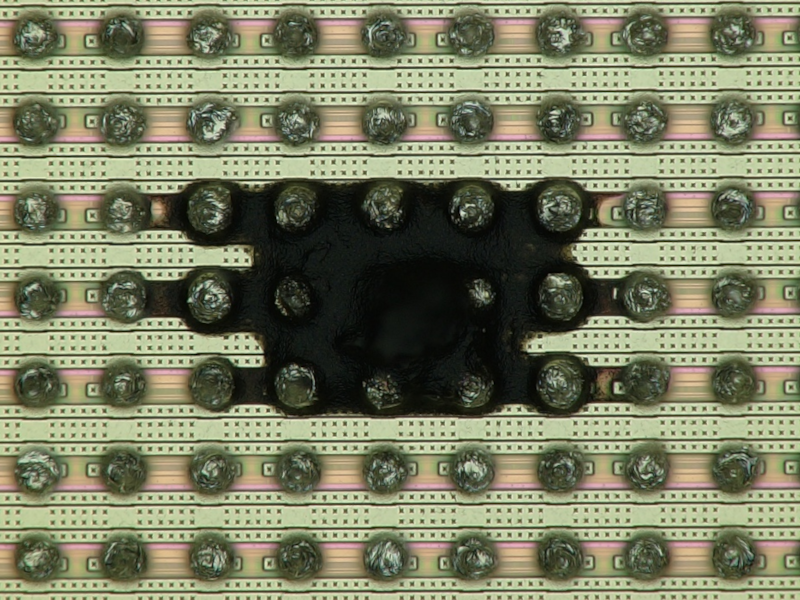}

\includegraphics[width=\textwidth]{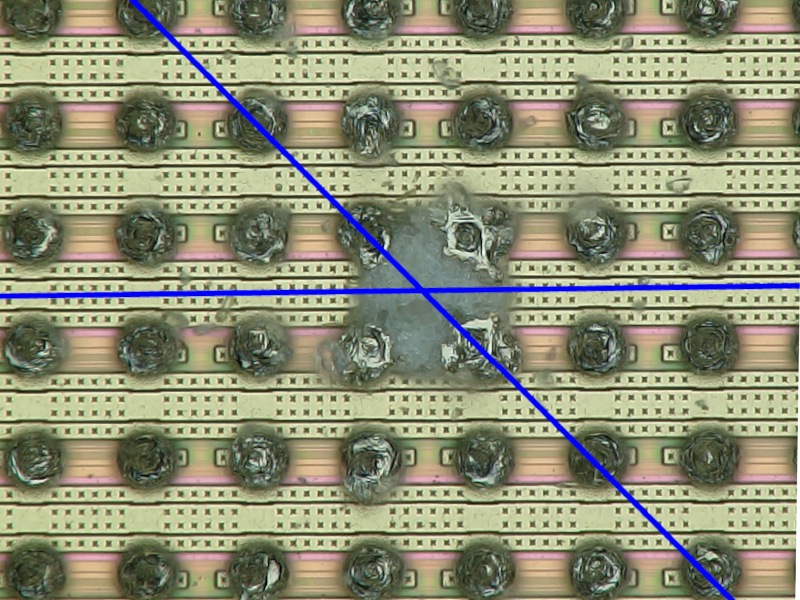}
\caption{}
\label{fig:birdview}
\end{subfigure}
\begin{subfigure}[b]{0.29\textwidth}
\includegraphics[width=\textwidth]{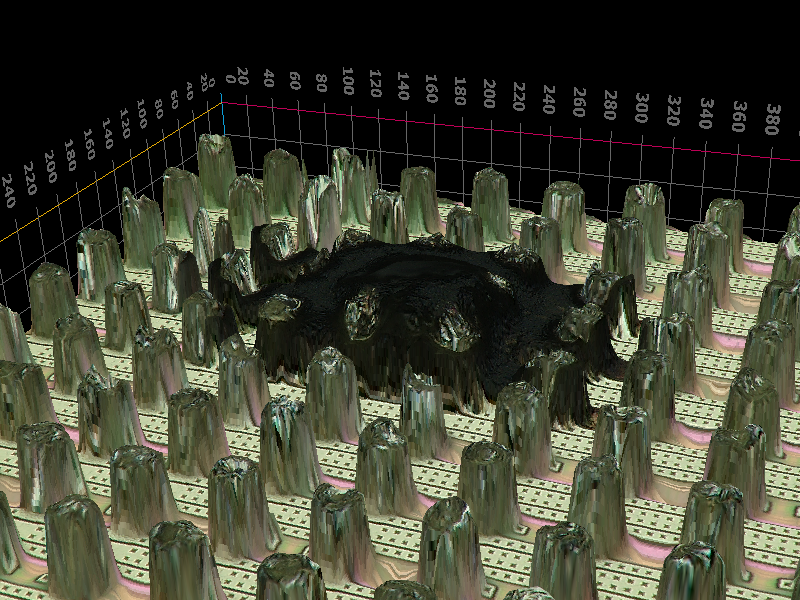}

\includegraphics[width=\textwidth]{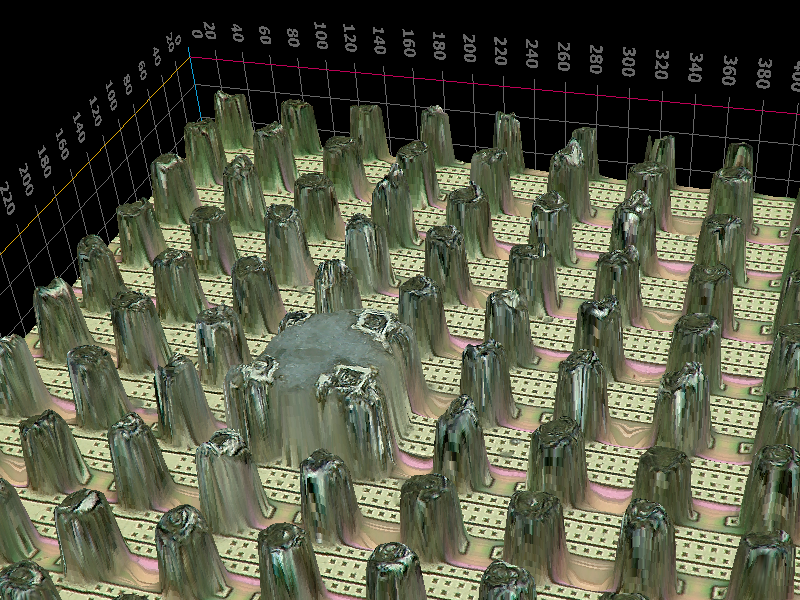}
\caption{}
\label{fig:3d}
\end{subfigure}
\begin{subfigure}[b]{0.39\textwidth}
\includegraphics[width=\textwidth]{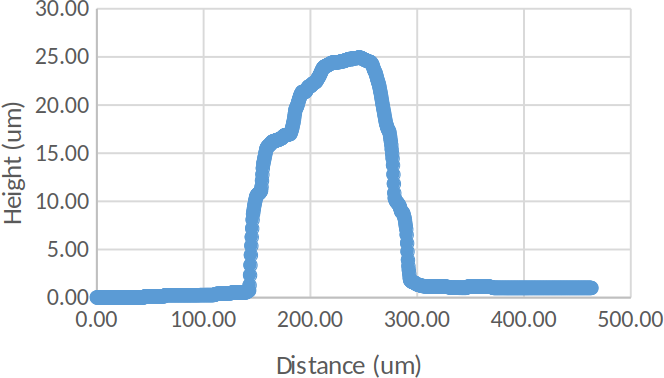}

\includegraphics[width=\textwidth]{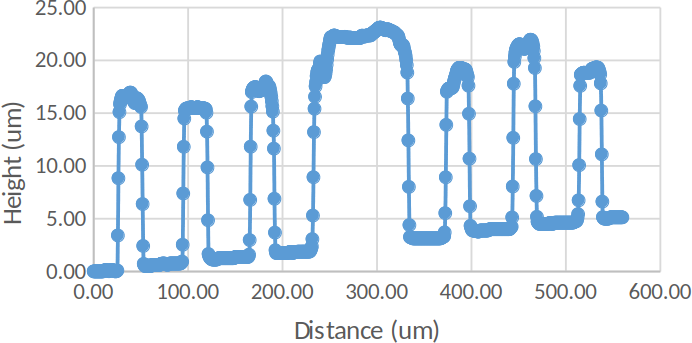}
\caption{}
\label{fig:height}
\end{subfigure}

\caption{Top: defect 1 does not correspond to a disabled \ac{cc}. Bottom: defect 3 is associated with a disabled \ac{cc} (same appearance as on the second disassembled \ac{cc} module). (a) birds-eye view of the defect. (b) 3D view of the defect. (c) measured dimensions of the defect.}
\label{fig:debris}
\end{figure}

Two modules with \ac{cc} issues from the same vendor were disassembled by heating up on a hotplate, shown in figure \ref{fig:hotplate}, and using vacuum tools to pull apart the chips at the melted solder bumps.
Three defects were found on one of the chips, shown as an example in figure \ref{fig:chip3defects}, where defect 1 and 3 are similar as can be seen in figure \ref{fig:debris}.
The location of disabled \acp{cc} agrees with the location of defect 3.
The appearence of the defect correlated with the \ac{cc} issue on the second module is the same as defect 3.
An \ac{edx} measurement revealed that the extra material consists mostly of silicon, which may be due to some debris from dicing that got between the \ac{fe} chip and the sensor during hybridisation.

\section{Conclusion}

The core column issue observed on pixel modules affects about 20--25\% of ITkPix v1.1 and v2 modules.
Our investigations uncovered inconsistent symptoms in the behaviour of the chip, showed potential dependency with chip's analog settings and confirmed the cause being mechanical damage as we first suspected.
While we are investigating deployment of enhanced \ac{qc} procedures at hybridisation vendors, we also prepare to check for bad core columns within our module \ac{qc} procedure by using the software tools developed.
However, due to the origin of the issue, the development throughout the whole \ac{qc} process and longterm impact are not yet known, especially regarding handling, thermal stress and irradiation.
Impact on tracking performance is being studied via simulations.


\begin{thebibliography}{99}


\bibitem{atlas}
ATLAS Collaboration, \emph{The ATLAS Experiment at the CERN Large Hadron Collider}, JINST 3 (2008) S08003

\bibitem{itk}
C. Buttar for the ATLAS ITk Collaboration, \emph{ATLAS ITk pixel detector overview}, \doi{10.1016/j.nima.2024.169978}

\bibitem{rd53}
RD53 Collaboration, \emph{RD53 pixel readout integrated circuits for ATLAS and CMS HL-LHC upgrades}, JINST 20 (2025) P03024



\bibitem{pdb}
M. Wielers for the ATLAS Collaboration, \emph{ATLAS ITk Production Database use and tools}, \href{https://cds.cern.ch/record/2923886}{ATL-ITK-PROC-2025-006}, CERN, Geneva (2024).


\end{thebibliography}
\end{document}